\documentclass[letter]{aa} 

\authorrunning{Orozco Su\'arez et al.\/} 
\titlerunning{Magnetic field emergence in quiet Sun granules} 

\usepackage{graphicx}
\usepackage{txfonts}
\usepackage{color}

\begin{document}
   \title{Magnetic field emergence in quiet Sun granules}

   \author{D.\ Orozco Su\'arez \inst{1}, L.R.\ Bellot Rubio \inst{1}, 
   J.C.\ del Toro Iniesta\inst{1} \and S.\ Tsuneta\inst{2}}

   \institute{Instituto de Astrof\'{\i}sica de Andaluc\'{\i}a (CSIC),
Apdo.\ Correos 3004, 18080 Granada, Spain \\
 \email{orozco@iaa.es} \and
National Astronomical Observatory of Japan, 2-21-1 Osawa, Mitaka,
Tokyo 181-8588, Japan}

   \date{Received ; accepted }

  \abstract
   {}
{We describe a new form of small-scale magnetic flux emergence in 
the quiet Sun. This process seems to take vertical magnetic 
fields from the solar interior to the photosphere, where they 
appear above granular convection cells.}
{High-cadence time series of spectropolarimetric measurements 
obtained by \emph{Hinode} in a quiet region near disk center are 
analyzed. We extract line parameters from the observed Stokes 
profiles and study their evolution with time. }  
{The circular polarization maps derived from the observed \ion{Fe}{i}
630~nm lines show clear magnetic signals emerging at the center of
granular cells. We do not find any evidence for linear polarization 
signals associated with these events. The magnetic flux patches grow 
with time, occupying a significant fraction of the granular area. 
The signals then fade until they disappear completely. The typical
lifetime of these events is of the order of 20 minutes. No significant
changes in the chromosphere seem to occur in response to the emergence, 
as revealed by co-spatial \ion{Ca}{ii} H filtergrams. The Stokes $I$ 
and $V$ profiles measured in the emerging flux concentrations show 
strong asymmetries and Doppler shifts.}
{The origin of these events is unclear at present, but we suggest 
that they may represent the emergence of vertical fields lines from 
the bottom of the photosphere, possibly dragged by the convective
upflows of granules. Preliminary inversions of the Stokes 
spectra indicate that this scenario is compatible with the 
observations. The emergence of vertical field lines is not 
free from conceptual problems, though.}
 {}

\keywords{Sun: magnetic fields -- Sun: photosphere 
-- Instrumentation: high angular resolution}

\maketitle

%

\section{Introduction}
  \label{sec:intro}

Our knowledge of the quiet Sun magnetism has improved dramatically in
the last years. We now have a relatively good understanding of the
properties of the fields and other aspects such as the small-scale
emergence and disappearance of magnetic flux concentrations. Analyses
of spectropolarimetric data with moderate spatial resolution have
significantly contributed to these topics.  Lites et al.\ (1996), for
example, discovered transient small-scale horizontal fields in quiet
Sun areas. More recently, Mart\'{i}nez Gonz\'{a}lez et al.\ (2007)
have presented convincing evidence of low-lying loops connecting
opposite-polarity flux concentrations in the solar internetwork.
Using MDI longitudinal magnetograms at 1\farcs2 resolution, Lamb et
al.\ (2007) studied the emergence of apparently unipolar flux in the
photosphere. Similar events have been detected in high-spatial
resolution magnetograms obtained at the Swedish Vacuum Telescope (de
Pontieu 2002).

Improvements in the spatial resolution of spectropolarimetric
measurements allow these processes to be studied in greater detail.
Using the spectropolarimeter aboard {\em Hinode}, Lites et al.\
(2007a,b) and Orozco Su\'arez et al.\ (2007a,b) have found that
internetwork fields tend to be horizontal with strengths below
0.5~kG. Also based on {\em Hinode} data, Centeno et al.\ (2007) and
Ishikawa et al.\ (2007) have confirmed the existence of very
small-scale magnetic loops in quiet Sun areas and plage regions,
respectively. The loops show horizontal fields above granules and
footpoints of opposite polarity rooted in the adjacent intergranular
lanes. The frequency of appearance of these loops seems to be much
higher than ever thought.

The characterization of emergence processes at the smallest scales is
important to understand the energy balance and origin of quiet Sun
magnetic fields. In particular, they may hold the key to determine
whether a local dynamo operates in the solar photosphere, as has been
suggested on theoretical grounds (Cattaneo 1999). Also, a good
knowledge of the ways magnetic fields emerge in the surface may help
refine numerical simulations of magnetoconvection such as those
performed by V\"ogler et al.\ (2005), Schaffenberger et al.\ (2006),
Stein \& Nordlund (2006), and Abbett (2007). If some form of flux
emergence is not observed in the simulations, additional ingredients
might need to be implemented in current codes.

Here we analyze spectropolarimetric measurements of the quiet Sun
taken with \emph{Hinode} at a resolution of 0\farcs32. Several
isolated, apparently unipolar, vertical field emergence events 
have been observed to occur in granules. We describe them and 
speculate on their origin in the next sections.

\begin{figure*}[!t]
\centering 
\includegraphics[height=7.8cm]{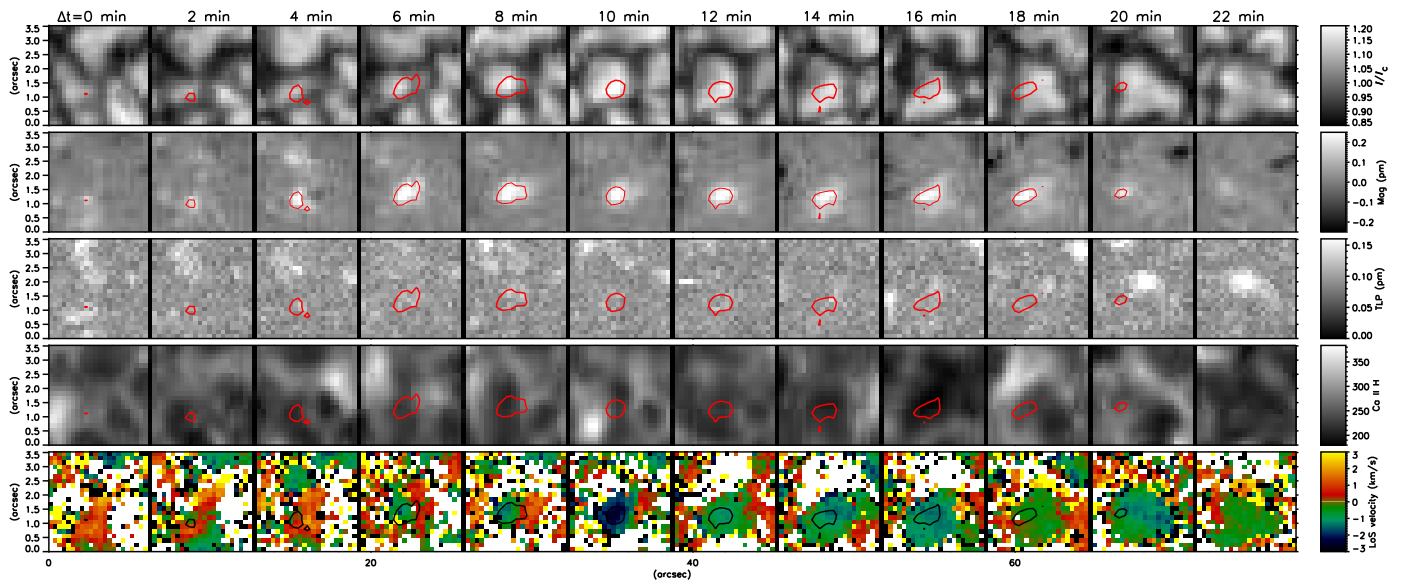}
 
\caption{Temporal evolution of a flux emergence process observed on
February 11, 2007. The cadence is 123~s and the maps cover a FOV of
$3\farcs5 \times 3\farcs5$. {\em From top to bottom:} normalized
continuum intensity, circular polarization signal in the blue lobe of
\ion{Fe}{i} 630.25~nm (integrated over a wavelength range of 23 pm
centered on the peak position), total linear polarization signal
[$\int{(Q^2+U^2)^{1/2}\mathrm{d}\lambda} / I_\mathrm{c}$],
\ion{Ca}{ii}~H line core intensity, and LOS velocity evaluated from
the Stokes $V$ zero-crossing wavelengths. Negative velocities indicate
blueshifts. The contours enclose areas with circular polarization
signals larger than 0.15~pm. White areas represent pixels not included in
the analysis because of their small signals. $\Delta t=0$ min
corresponds to 11:07 UT}

\label{fig:fig1}
\end{figure*}

\section{Observations}
\label{sec:obser}

The data under analysis consists of two sequences of raster scans
performed with the spectropolarimeter (SP; Lites et al.\ 2001) aboard
{\em Hinode} (Kosugi et al.\ 2007). This instrument measures the
Stokes profiles of the \ion{Fe}{i} lines at 630.2~nm with a spectral
sampling of 2.15~pm pixel$^{-1}$ and a spatial sampling of
0\farcs16. The sequences were taken on February 11 and March 10,
2007. In both cases, a narrow $4\arcsec \times 82\arcsec$ area at disk
center was scanned repeatedly with a cadence of 123 s. The integration
time for each of the 25 slit positions was 4.8~s, making it possible
to achieve a noise level of $1.1 \times 10^{-3} \, I_{\rm c}$ in
Stokes $V$ and $1.2\times10^{-3} \, I_{\rm c}$ in Stokes $Q$ and $U$.
Filtergrams in the \ion{Ca}{ii}~H line core were also acquired with
the {\em Hinode} Broadband Filter Imager (BFI; Tarbell et al.\ 2007)
to monitor the conditions of the chromosphere. The BFI pixel size was
0\farcs054 on February 11 and 0\farcs108 on March 10, and the cadences
were 64 and 32 s, respectively. Given that magnetic fields outside the
photospheric network are not associated with enhanced \ion{Ca}{ii}~K
or H emission (Lites et al.\ 1999; Rezaei et al.\ 2007), the detection
of transient \ion{Ca}{ii}~H brightenings at or near the position of
emerging flux would indicate that the emergence process is able to
transfer a certain amount of energy to the (low) chromosphere.

The SP data have been corrected for dark current, flat-field, and
instrumental cross-talk. The velocity scale for the Stokes spectra has
been set by comparing the average quiet-Sun profile in each map with
the FTS atlas, after subtraction of the gravitational redshift. The
calibration algorithm applied to the filtergrams removed cosmic rays, 
hot pixels, and dark current. SP maps and BFI filtergrams have been
aligned with sub-pixel accuracy following Shimizu et al.\ (2007).

%

\section{Results}

\begin{figure*}
\centering 
\includegraphics[height=7.9cm]{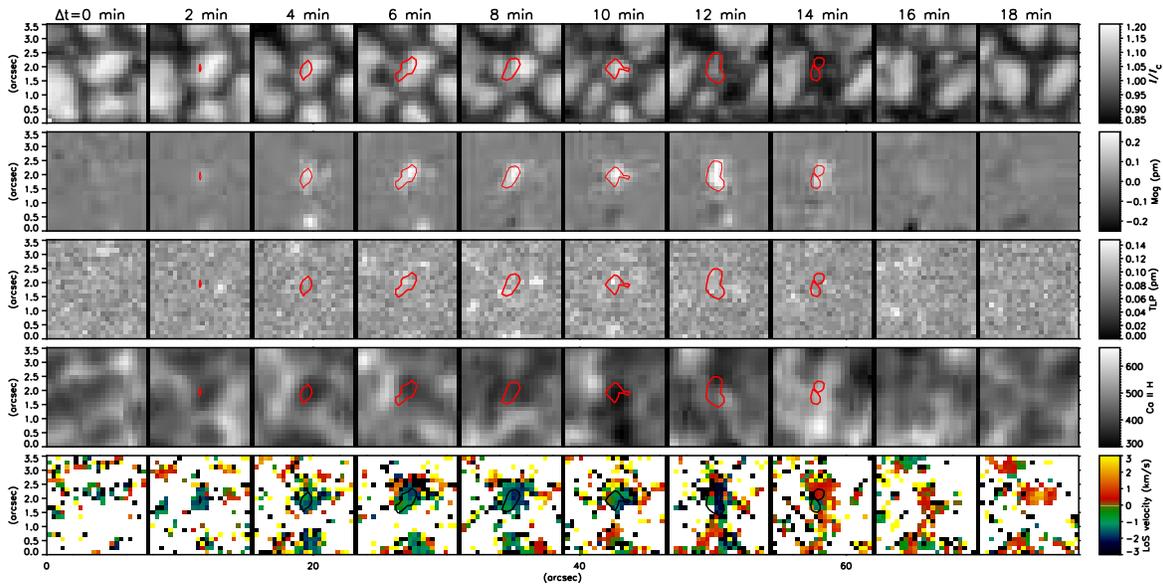} 
\caption{Emergence event observed on March 10, 2007. The panels
show the same quantities as those of Fig.~\ref{fig:fig1}. $\Delta t=0$ 
corresponds to 00:52 UT.}
\label{fig:fig1_bis}
\end{figure*}

Figures~\ref{fig:fig1} and \ref{fig:fig1_bis} show the temporal
evolution of two small-scale emergence events. Displayed are maps 
of continuum intensity at 630~nm, circular and linear polarization
signals, \ion{Ca}{ii} filtergrams, and LOS velocity maps. The total
duration of the events is about 20 and 14 minutes, respectively.

In the first event, the circular polarization maps show a unipolar
flux concentration (white patch) barely visible at $\Delta
t=0$~min. As time goes by, it grows both in size and strength. The
maximum size and circular polarization signals are reached 8 min
later. At that point, the flux concentration looks roundish and
occupies an area of $\sim 4\times 4$ pixels (some 200\,000 km$^2$; red
contours), which corresponds to one third of the granular cell
surface. The granule is defined to be the region where the continuum
intensity is at least 1.05 times brighter than the average quiet
Sun. Then the signal starts to fade away (16 min). At $\Delta t = 24$
min (not shown), the circular polarization signal has vanished
completely.  No clear opposite-polarity signals are detected in the
observed area during the whole sequence. The continuum intensity maps
demonstrate that the magnetic flux appears in an existing granule and
persists there for 20 min while the granule evolves. Interestingly,
the flux concentration does not seem to be disturbed by the granular
flows: it remains co-spatial with the brightest part of the granule
until $\Delta t=16$~min and never gets advected to the adjacent
intergranular lanes.

There is no detectable linear polarization signal associated with this
event. Only the last two maps show traces of linear polarization when
the circular polarization signal is almost absent. The observed linear
polarization patch lies close to the flux concentration (less than
1\arcsec\/ up and right), but we believe it is not related to its
disappearance.

The emergence is characterized by blueshifted velocities already from
the initial stages. At $\Delta t=10$ min we observe the stronger
upflows of about $-2.5$~km~s$^{-1}$. At that moment, a chromospheric
brightening is detected in the \ion{Ca}{ii}~H filtergrams. It might be
related to the emergence process, although it is not co-spatial with
the flux concentration. 

The second event differs slightly from the previous one. The magnetic
feature appears at $\Delta t=2$~min and persists for 12 min before it
vanishes. The maximum spatial size of the magnetic patch is not larger
than in the previous case.  However, it occupies about half of the
granule at $\Delta t=6$ min, and almost the whole granule 4 min later.
The circular polarization signal is maximum at $\Delta t=14$ min. As
before, there is no evidence for horizontal magnetic fields since no
linear polarization is detected. The continuum intensity maps show
that the granule starts to diminish in size at $\Delta t = 10$ min,
disappearing completely four minutes later. The flux concentration
still persists, however.  Thus, our observation suggests that the
magnetic field somehow contributed to the granular disruption.
Throughout the sequence, the flux concentration appears to be
decoupled from the advection flow (e.g., it moves towards the center
of the granule between $\Delta t=6$ min and $\Delta t=10$~min).

Similarly to the first case, blueshifts are observed right from the
beginning of the process. From $\Delta t = 16$ min on, weak downflows
are detected instead, corresponding to the granule disappearance. The
largest upflows of $-2.2$~km\,s$^{-1}$ are comparable with those of
the first case. The \ion{Ca}{ii}~H line shows no significant 
brightenings associated with the process.

\begin{figure}  
\centering 
\includegraphics[width=6.9cm]{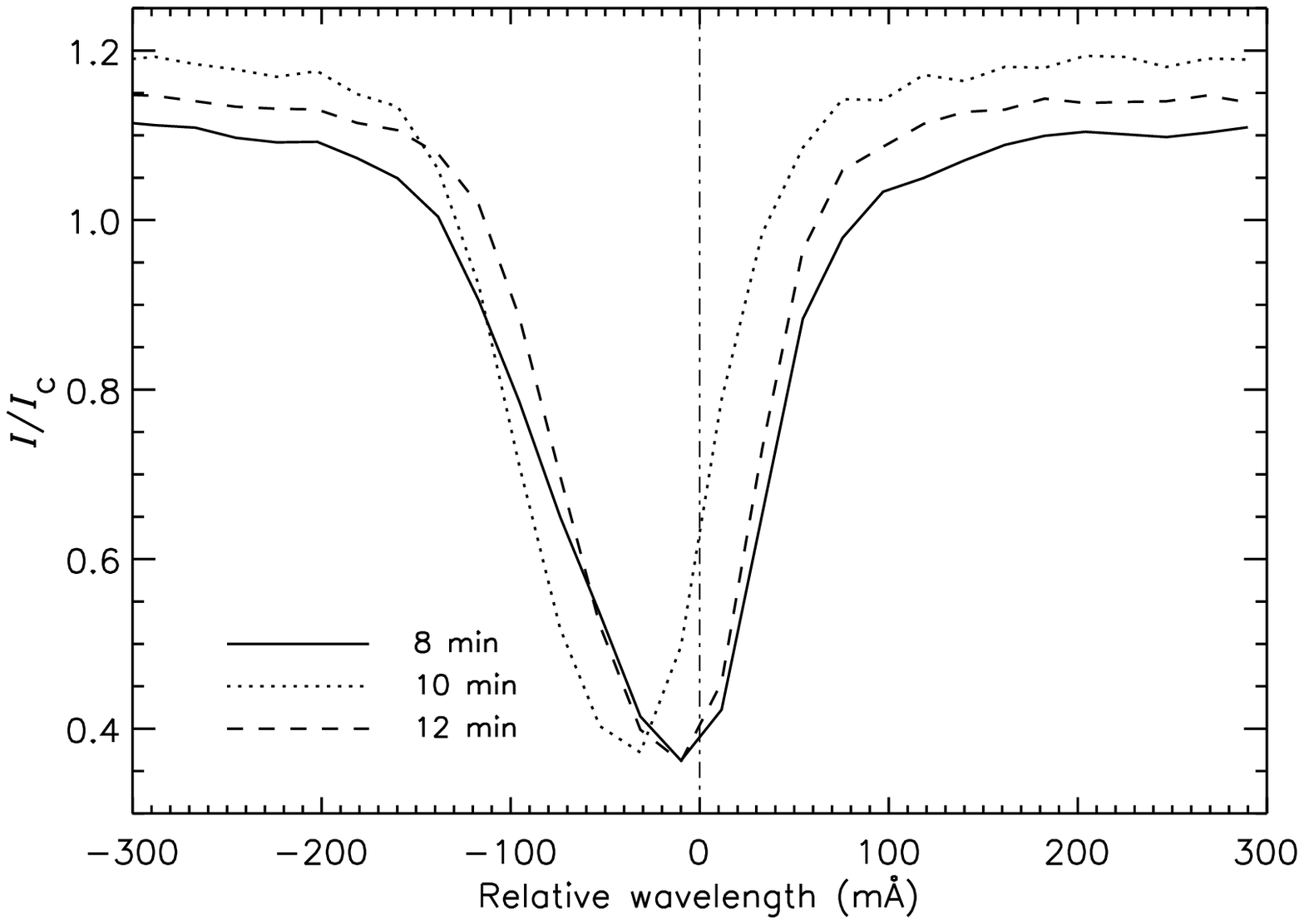}
\includegraphics[width=6.9cm]{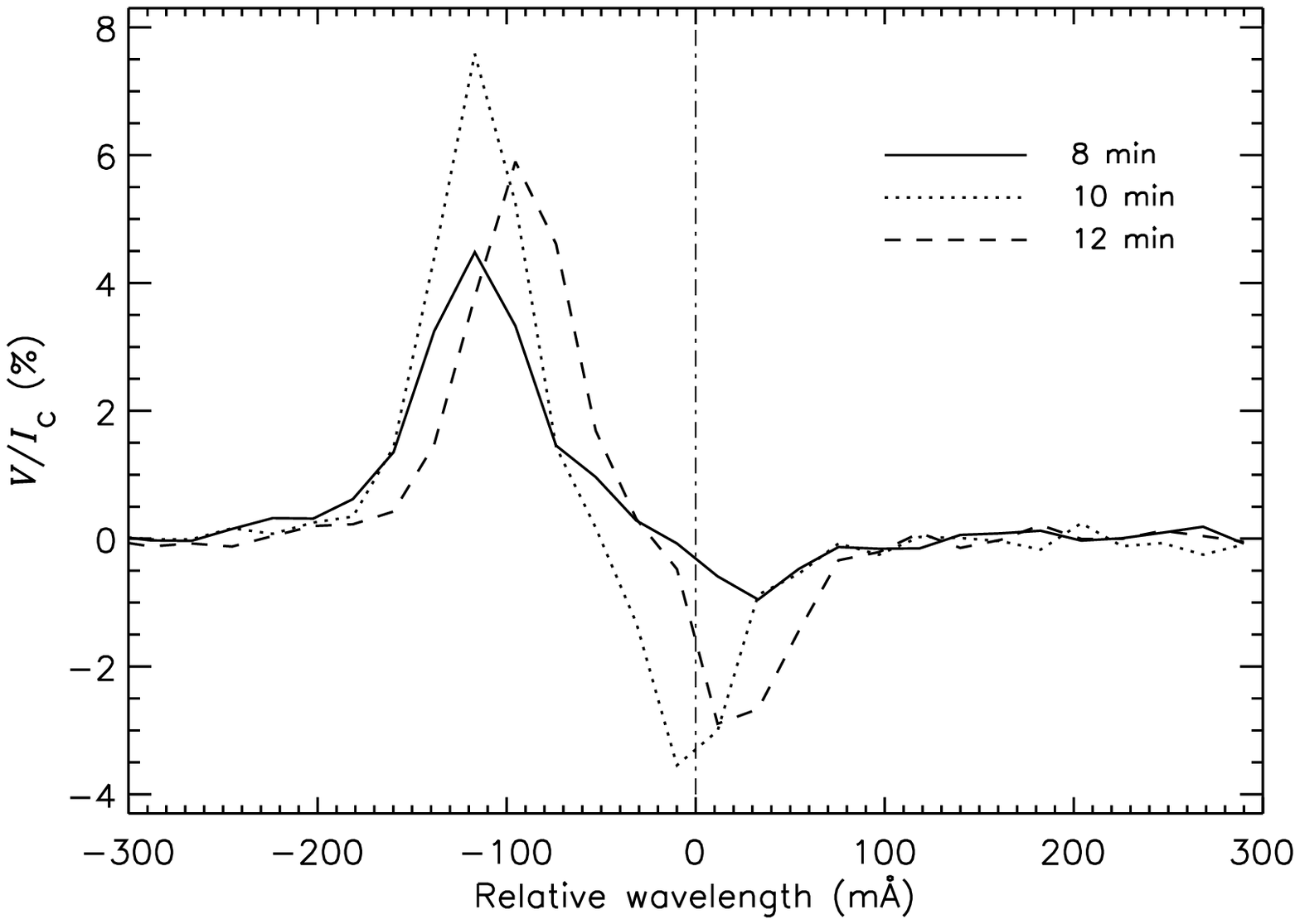} 
\caption{Stokes $I$ (\emph{top}) and $V$ (\emph{bottom}) profiles of
\ion{Fe}{i} 630.25~nm observed at the center of the flux concentration
studied in Fig.~\ref{fig:fig1}. The solid, dotted and dashed lines
correspond to $\Delta t = 8$, 10, and 12 minutes, respectively.  The
vertical lines mark the zero point of the velocity scale.}
\label{fig:fig2}
\end{figure}

Figure~\ref{fig:fig2} displays Stokes $I$ and $V$ profiles observed
during the first event. They correspond to the center of the flux
concentration at $\Delta t = 8$, $10$ and $12$ minutes (solid, dotted,
and dashed lines, respectively). We do not include Stokes $Q$ and $U$
spectra because their signals are below the noise.

Both Stokes $I$ and $V$ exhibit strong asymmetries at $\Delta t = 8$
min. The blue wing of Stokes $I$ is significantly blueshifted while
the line core remains almost at rest, suggesting strong upflows in
deep atmospheric layers and smaller velocities higher up. The
signature of large gradients of atmospheric parameters is even more
conspicuous in Stokes $V$: since the pioneering work by Illing et al.\
(1975) and Auer \& Heasley (1978) it is known that the circular
polarization profiles are symmetric unless a velocity gradient is
present along the LOS. The asymmetry of Stokes $V$ is extreme in this
case, with the red lobe being almost absent. Such a degree of
asymmetry can only be produced by large velocity {\em and} magnetic
field gradients. At $\Delta t = 10$ min, the whole line is affected by
a strong blueshift, but the velocity gradient seems to have decreased
significantly since the profiles look more symmetric.  At $\Delta t =
12$ min, the gradients are still small and the global velocity shift
is reduced. Altogether, this qualitative interpretation of the Stokes
$I$ and $V$ profiles suggests that we are witnessing the rise of
magnetic fields through the granule, from the bottom of the
photosphere to higher layers.  Apparently, the field is vertical
because we do not detect linear polarization signals.

\section{Preliminary inversions}

We have inverted the profiles displayed in Fig.~\ref{fig:fig2} to
estimate the field strength and flow velocity in the emerging magnetic
flux concentration. To that end we have used SIRJUMP, a modified
version of the SIR code (Ruiz Cobo \& del Toro Iniesta 1992). Unlike its
parent, SIRJUMP is able to deal with discontinuous stratifications
along the LOS. The discontinuity is modeled as a step function. Tests
with the code show that it successfully distinguishes cases where a
discontinuity is needed from those that do not require it. We run
SIRJUMP with one-component model atmospheres and unity filling
factors, allowing for a local stray light contamination as described
by Orozco Su\'arez et al.\ (2007a,b).

The results of these {\em preliminary} inversions indicate that the
only way to reproduce the Stokes $I$ and $V$ profiles observed at
$\Delta t = 8$ min is by means of a discontinuous stratification of
magnetic field strength. A weak vertical field of about 250~G is
present in the lower photosphere, together with strong upflows of some
$-2.5$~km\,s$^{-1}$.  The field strength is negligible above $\log
\tau_5 \sim -1$. The discontinuity moves to higher layers at $\Delta t
= 10$ min and eventually leaves the line-forming region at $\Delta t =
12$ min. The scenario coming out from the inversions is therefore
compatible with the qualitative suggestions made from the shapes of
the profiles: in the first stages of the emergence we could be seeing
the ascent of magnetic fields that do not fill the line-forming region
completely.  Once the fields reach the upper photosphere, the
gradients of atmospheric parameters decrease significantly and, 
as a consequence, the Stokes profiles become less asymmetric.


  \section{Discussion and conclusions}
  \label{sec:con}

In this Letter we have described two cases of the emergence of
apparently unipolar, vertical fields in granular cells. A total 
of six such events occurred during the 10 hours covered by the 
Hinode observations. Their lifetimes are of the order of 15-20 minutes.

To the best of our knowledge, this form of small-scale magnetic flux
emergence has not been described in the literature.  It differs
significantly from the emergence processes in granular convection
studied by Centeno et al.\ (2007) and Ishikawa et al.\ (2007), since
we do not detect linear polarization signals or opposite-polarity
footpoints surrounding them. The geometry of the fields we observe is
not that of small magnetic loops. Lamb et al.\ (2007) have described
examples of the emergence of unipolar flux, but at a poorer resolution
of 1\farcs2.  No association of the flux with granules or 
intergranules was made in their paper. They suggested that the origin
of the unipolar flux appearance is coalescence of pre-existing field
lines with the same polarity, which were below the detection limit
imposed by the intrinsic noise and the spatial resolution of their
observations. While it is not possible for us to rule out the scenario
of field-line coalescence, we do not find evidence for diffuse
magnetic fields in the emergence sites prior to the events, at the
much higher spatial resolution and sensitivity of Hinode.

Current magnetoconvection simulations do not seem to explain our
observations either.  The simulations of V\"ogler et al.\ (2005) do
show magnetic fields in granules, but they are much weaker than those
reported here, and do not undergo emergence processes. Those granular
fields may be the result of recycling of flux initially placed in
intergranular lanes, or an effect of enhanced magnetic diffusivities.
Cheung et al.\ (2007), on the other hand, have studied the rise of
magnetic flux tubes from the convection zone to the photosphere. 
Depending on the magnetic flux stored in the tubes, the arrival 
of magnetic fields at the solar surface has very different 
observable consequences.  For the stronger tubes, a darkening and
distortion of the granular convection is expected (and actually
observed), while weaker tubes do not modify the brightness of surface
granules. In both cases, magnetic fields tend to emerge at the 
center of granular cells, showing large inclinations to the vertical. 
The fields are then advected by the horizontal flow towards the 
intergranular lanes, where they become more vertical and form 
opposite-polarity patches. Our events do not share these 
properties. 

A hypothetical scenario for the emergence of unipolar vertical
magnetic fields would be that granular upflows drag horizontal field
lines initially placed in the upper convection zone, carrying them to
the photosphere where they would emerge in the granules. However, it
is not clear how the horizontal fields may turn into vertical
fields. Also, at some point one should observe opposite polarities
where the field lines return to the solar surface, but we do not
detect them, perhaps as a consequence of still insufficient
sensitivity or because they occur outside of the FOV. What is clear is
that the scenario of vertical fields emerging in granules faces
important conceptual problems.

A radically different interpretation is that the events we have
observed do not involve the emergence of new flux, but the
``excitation'' of already existing, mixed, quasy-isotropic fields
(L\'opez Ariste et al.\ 2008).  These fields would be largely
decoupled from convective motions and hence not affected by them.  If
the degree of mixing is sufficiently high, the absence of linear
polarization cannot be taken as a proof that the field is
vertical. This scenario should be investigated more thoroughly, both
to demonstrate the existence of such tangled fields and to assess
whether they are indeed compatible with the observations presented
here.

In summary, at this stage we cannot offer a clear explanation of 
the events observed with Hinode, but we hope that the report of 
such a process will stimulate further observational and theoretical
work. With this in mind, we plan to perform detailed inversions of the
observed profiles in an attempt to derive a consistent picture of the
physics behind these processes.  Since sensitivity may be an issue, it
would also be convenient to carry out additional measurements with the
{\em Hinode} SP, pushing the integration time to a limit.


\acknowledgements {\em Hinode} is a Japanese mission developed and launched
by ISAS/JAXA, with NAOJ as domestic partner and NASA and STFC (UK) as
international partners. It is operated by these agencies in
co-operation with ESA and NSC (Norway). This work has been partially
funded by the Spanish Mi\-nisterio de Educaci\'on y Ciencia through
project ESP2006-13030-C06-02 (in\-clu\-ding European FEDER funds).

\end{document}